\documentclass[iop]{emulateapj}
\usepackage{url}
\usepackage{graphicx}
\usepackage{xcolor}
\usepackage{soul} 

\usepackage{color}
\usepackage{amsmath}
\usepackage{setspace}
\usepackage{threeparttable}
\usepackage{epstopdf}
\usepackage{graphics}
\usepackage{subfigure}
\usepackage{url}             
\usepackage{rotating}
\usepackage{lipsum}
\usepackage{float}
\usepackage{wrapfig}
\usepackage{longtable}
\usepackage{natbib}
\usepackage{hyperref}

\usepackage{enumitem}

\graphicspath{{./}{figures/}}

\begin{document}
\nocite{*}
\title{ A QPO in NGC 4945 from Archival $\it{RXTE}$ Data}

\author{Evan Smith, Rebecca Robles, Eric Perlman}
\affiliation{Aerospace, Physics and Space Sciences Department, Florida Institute of Technology, 150 West University Blvd., Melbourne, FL. 32901, USA }



\begin{abstract} 
We report the discovery of a $\sim$6-week quasi-periodic oscillation (QPO) in archival NGC 4945 data observed by the $\it{Rossi~X}$-$\it{ray~Timing~Explorer~(RXTE)}$ satellite. QPOs are an important observable in accretion disks and have been studied extensively in both neutron star (NS) and black hole (BH) X-ray binaries (XRB). QPOs should be present in Active Galactic Nuclei (AGN) if galactic black holes and supermassive black holes (SMBH) are governed by a common set of physical processes. The search for QPOs in AGN has proven difficult because the timescales would be much longer, due to their higher mass. $\it{RXTE}$ AGN light curves spanning 1996 to 2011 provide an excellent and perhaps unique opportunity to search for low-frequency QPOs. We investigated the 533 $\it{RXTE}$ observations made of the Seyfert-2 AGN, NGC 4945. During a large cluster of observations in 2006-2007 (194 observations, spanning 396 days), the Lomb-Scargle periodogram shows a candidate QPO at 0.274 $\mu$Hz (period $\approx$ 42.2 days). We estimate the uncertainties using the False Alarm Probability (FAP). We discuss the possible identification of this feature with the Lense-Thirring precession period.
\end{abstract}

\section{Introduction}
Quasi-Periodic Oscillations (QPOs) are an important observable in accretion disks. QPOs have been studied extensively in both neutron star and black hole X-ray binaries (BHXRB). In galactic sources, kilohertz (kHz) or high-frequency (HF) QPOs are believed to probe the motion of matter in strong gravity, within a few kilometers of the NS surface or BH event horizon. There is every indication that we are observing orbits limited by the innermost stable circular orbit (ISCO) predicted by general relativity, and Lense-Thirring precession \citep{1918PhyZ...19..156L} of these orbits caused by frame dragging associated with the central object's spin.

The history of QPO models is traced in reviews past \citep{1989ARA&A..27..517V, 2000ARA&A..38..717V} and current \citep{2020arXiv200108758I}. Possible explanations for QPOs include Keplerian orbital motion of matter in the disk, spin of the central compact object, general relativistic effects, or beat frequencies between two of the previous mechanisms. A common element in most QPO models is a sonic radius, the disk radius where the radial inflow velocity becomes supersonic. RXTE observations of kilohertz burst oscillations were used to constrain NS spin periods \citep{1999ApJ...516L..81S}.

$\it{RXTE}$ observations of microquasars revealed QPOs at nearly identical frequencies: 67 Hz for GRS 1915+105 and 66 Hz for IGR J17091-3624. Common families of complex light curves in these two sources prompted investigation of whether the behaviors are produced by the same physical mechanisms \citep{2011ApJ...742L..17A}. A fundamental question in astrophysics is whether galactic black holes in BHXRB and SMBH in AGN are governed by a common set of physical processes. The disk-jet connection was investigated \citep{2003MNRAS.345.1057M} finding tight correlations between BH mass, radio and x-ray luminosity, over many orders of magnitude. This ‘Fundamental Plane’ supports the theory of scale invariance, suggesting a commonality in the underlying physics of disk-accretion and jet-launching in BHXRB and AGN. Thus, observational evidence of timing features would appear on much different time scales, which are often compared using the mass ratio.

The ISCO for matter in the accretion disk around a black hole has radius $R_{ISCO} \equiv 3 R_{S} \equiv \frac{6 G M}{c^2}$ for a non-spinning or Schwarschild black hole. For the rotating or Kerr black hole, $\frac{G M}{c^2} \leq R_{ISCO} \leq \frac{9 G M}{c^2}$, where the lower and upper limits are for maximally-rotating BH in the co-rotating and counter-rotating cases, respectively \citep{2015PhRvD..91l4030J}. \textbf{For a steady-state disk model \citep{1973A&A....24..337S}}, the disk temperature at ISCO, $T_{ISCO}$, varies as $M^{-1/4}$ \textbf{and temperature varies with radius as $r^{-3/4}$}. For AGN, $T_{ISCO}$ corresponds to thermal emission in the ultraviolet. The generally accepted model to explain detection of X-rays in these sources is inverse Compton scattering by electrons in a hot corona of about 100 keV. The effects of absorption and reflection on the X-ray spectrum are varied \citep{2009A&ARv..17...47T}, but it is generally accepted that the seed photons emanate from the accretion disk.

Recent reverberation mapping results \citep{2019Natur.565..198K} suggest that the corona is considerably smaller during the luminous hard state than in the thermal soft state. SMBH X-ray emission from a compact central corona can be sorted from corresponding reflection from the accretion flow as these have different spectral shapes. High-frequency reverberation time lags now allow study within a few gravitational radii of the event horizon. Techniques, modeling and instrumentation for these studies all show continual advancement \citep{2014A&ARv..22...72U}.

\begin{figure*}[t]
    \centering
    \includegraphics[scale=3, width=0.49\linewidth]{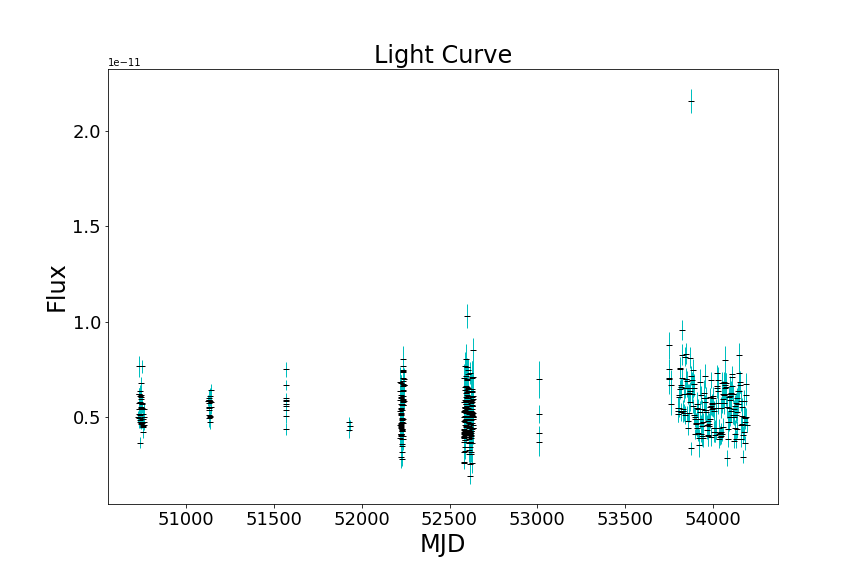}
%
    \includegraphics[scale=3, width=0.49\linewidth]{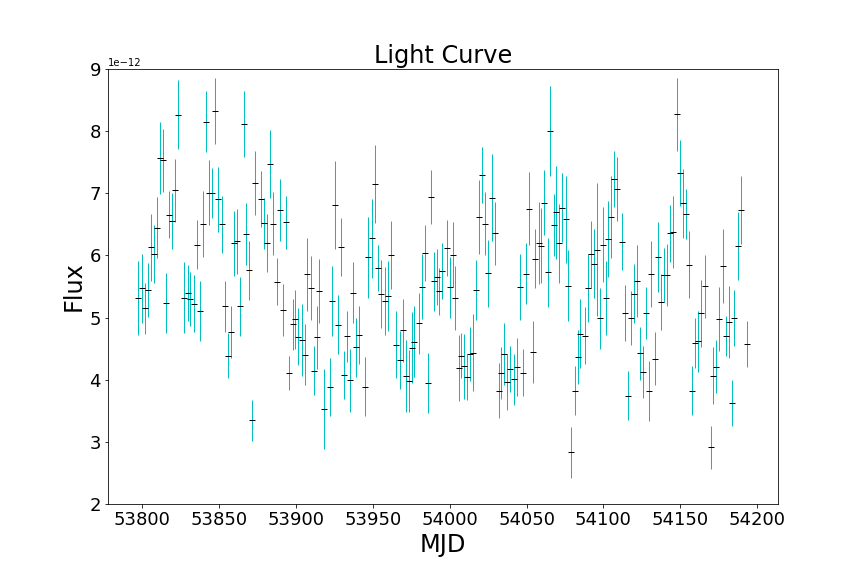}

       \caption{(left) The NGC 4945 light curve of all $\it{RXTE}$ observations (1997-2007). Notable clusters of observations are found spanning November and December 2002 (183 observations) and January 2006 to April 2007 (200 observations). The remaining observation clusters (42 in 1997, 18 in 1998, 10 in 2000, 3 in January 2001, 73 in November 2001, and 4 in 2004) were too brief for meaningful analysis, and the large gaps between clusters precluded combining segments in a single run. Two of the short clusters have 4 observations per day, and would have been sensitive to QPOs down to a lower limit of about 12 hours.}

    \caption{(right) The $\it{RXTE}$ NGC 4945 light curve from January 2006 to April 2007 is shown on an expanded scale. The average spacing is 2 days. An outlier on 2006 May 20 was analyzed \citep{2011ApJS..193....3R}, providing possible galactic, extragalactic or spurious explanations. We re-scaled the figure to avoid this point and give better vertical resolution. Ultimately, we analyzed several intervals within this cluster, and better fits were obtained starting after MJD 54000. The fluctuations are clearly evident, and look periodic in this restricted interval, after the outlier point.}
    \label{fig:2}
\end{figure*}

The search for QPOs in AGN has proven difficult. One claim fell out of favor after being attributed to detector malfunction \citep{1993Natur.361..233P,1996ApJ...465..181T}. For another candidate QPO, attempts to confirm the finding with additional observations have been unsuccessful \citep{1998MNRAS.295L..20I, 2004MNRAS.347..411I}. The breakthrough AGN QPO candidate detection was reported in the narrow-line Seyfert 1 galaxy, REJ 1034+396, found in an uninterrupted 91-ks $\it{XMM}$-$\it{Newton}$ observation from 2007, binned at 100-s intervals \citep{2010MNRAS.403....9M}. The QPO frequency was 2.7 $\times$ $10^{-4}$ Hz, or a period of 3733 $\pm$ 140 s. Using scaling arguments to compare this QPO with the 67-Hz QPO in the BHXRB GRS1915+105 (assuming mass of 10 $M_{\odot}$) provides a mass estimate for REJ 1034+396 in the range $\sim$(1-4) $\times 10^6 M_{\odot}$. Interpreting either peak as a Keplerian orbit implies a radial distance in the range of 20-25 $R_{S}$. The QPO in REJ 1034+396 was not originally detected in subsequent observations, but a more exhaustive search \citep{2014MNRAS.445L..16A} found five more occurances of this QPO feature. A later study using older data, an uninterrupted 70-ks $\it{XMM}$-$\it{Newton}$ observation from 2005, detected a 0.15-mHz QPO ($\sim$2-hr) in another narrow-line Seyfert 1 galaxy, MS 2254.9-3712 \citep{2015MNRAS.449..467A}. An iron K$\alpha$ lag is seen, and a recent study \citep{2016MNRAS.462..511K} found iron K reverberation signatures in about half of the Seyfert galaxies in the $\it{XMM}$-$\it{Newton}$ archive with sufficient variability and duration. 

\section{Observations}

To look for low-frequency (LF) QPOs, we have investigated archival data from the $\it{Rossi~X}$-$\it{ray~Timing~Explorer}$ ($\it{RXTE}$) satellite collected between 1996 and 2011. AGN monitoring on a regular cadence was performed for many sources, and reduced 3-color light curves have been prepared and archived at University of California, San Diego (UCSD) \citep{2011ApJS..193....3R,2013ApJ...772..114R}. 

\begin{figure*}[t]
    \centering
    \includegraphics[scale=3, width=0.49\linewidth]{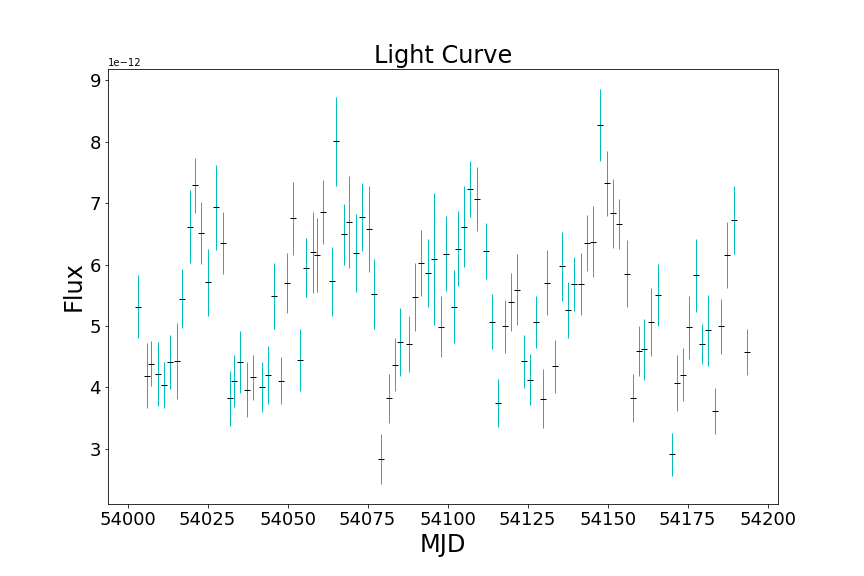}
%
    \includegraphics[scale=3, width=0.49\linewidth]{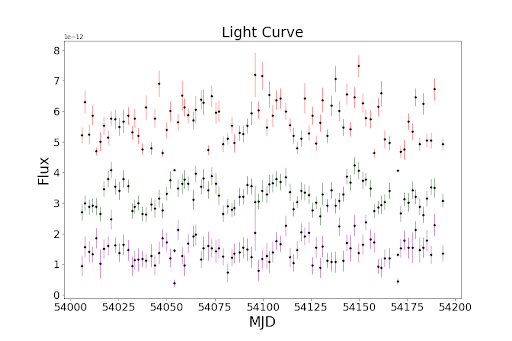}
   \caption{(left) $\it{RXTE}$ 2-10 keV Light Curve for NGC 4945, MJD 54003-54193. This is the interval for our primary result. The cycles are clearly visible in the light curve. Error bars on the individual observations were $< 10\%$.}

    \caption{(right) 3-Color Light Curve for NGC 4945; violet (7-10 keV), green (4-7 keV, +1$\times 10^{-12}$ ergs/$cm^2$/s), red (2-4 keV, +4$\times 10^{-12}$ ergs/$cm^2$/s). We analyzed the 3 sub-bands individually for timing features. The light curves are presented here with offsets so that they are separated enough to be viewed individually. A variation near the QPO frequency is evident in each sub-band, at much lower significance. The significance shows a modest increase with hardness.}
    \label{fig:4}
\end{figure*}
\subsection{The RXTE AGN Archive}
$\it{RXTE}$ conducted numerous AGN observations over its 16-year lifespan, on a total of 153 sources. The sampling strategies vary with different groups of proposers and for a variety of scientific aims. The strategy in assembling the $\it{RXTE}$ AGN archive at UCSD was simply to use all available data.  Of the AGN in the UCSD database, we found 76 with enough separate observations to support the search for low-frequency QPOs. The campaign lengths, as long as 15 years in some cases, provide time series sensitive to QPOs of longer duration than could be detected in other datasets. We selected AGN for study if they have both a large quantity of observations and appropriate spacing to support the search for QPOs. The NGC 4945 result reported here stood out in our initial reconnaissance of the $\it{RXTE}$ AGN archive \citep{2019HEAD...1710631S}. Other preliminary results, both positive and negative, have been reported \citep{2020AAS...23621201H,2020AAS...23621204R}. 

For the present analysis, each individual observation represents a single data point in the light curve, regardless of length. The archived value is the mean flux over the whole observation, and the time tag is the observation midpoint. Longer observations could be re-processed to give time series with much shorter bins, but this was not the primary aim. For each AGN observation, the $\it{RXTE}$ AGN archive at UCSD makes available the mean count rate (and its uncertainty) in the 2-10 keV spectral range, and for three sub-bands (2-4 keV, 4-7 keV and 7-10 keV). A total of 533 $\it{RXTE}$ observations were made of the Seyfert-2 AGN, NGC 4945. Many smaller clusters of observations are available between 1997 and 2004 (Figure 1), but we found the largest clustering in 2006-2007. There are 194 observations spanning 396 days (MJD 53797.4 - 54193.6; 2006 March 3 - 2007 April 3). During this interval (Figure 2), the spacing between observations varied from a minimum of 1.17 days to a maximum of 4.27 days. The mean spacing was 2.05 days, and the median spacing was 1.96 days.

\subsection{RXTE Data Reduction}
This analysis uses exclusively data from the $\it{RXTE}$ Proportional Counter Array (PCA) \citep{2006ApJS..163..401J}. Details of the data reduction can be found in the description of the UCSD archive \citep{2011ApJS..193....3R}. Essentially all AGN observations used the Guest Observer (GO) modes Good$\_$Xenon1$\_$2s and Good$\_$Xenon2$\_$2s. While 16s readout was also available, and AGN observations were very unlikely to saturate in 16 seconds, 2s readout allowed more flexibility to re-point to a brighter Target-of-Opportunity (TOO). 

A standard screening procedure was to exclude data within 20 minutes of RXTE passage through the South Atlantic Anomaly (SAA). Mission experience showed that high background was not experienced immediately prior to SAA entrance, but subsisted for well over 20 minutes after SAA exit. The PCA background model specifically includes contributions from electron events induced by the SAA, and was calibrated with dedicated sky background position data from a number of positions in regions believed to be free of bright X-ray sources. Most $\it{RXTE}$ orbits pass through the SAA, inducing background terms that decay exponentially. Orbits without SAA passage were assumed to be unaffected by previous passages, valid as long as the background decay timescale is shorter than an orbit. A residual term in the background is then modeled as flux induced by the SAA undergoing exponential decay with an e-folding time of 240 minutes (longer than the $\sim$96-minute $\it{RXTE}$ orbital period). The amplitude for this term is determined by integrating the rates from particle monitors on the $\it{RXTE}$ High Energy X-ray Timing Experiment (HEXTE). HEXTE particle monitor data have also been used for the purpose of mapping the long-term history and intensity of the SAA \citep{2009E&PSL.281..125F}.

\begin{figure}[t]
    \centering
    \includegraphics[scale=6, width=0.98\linewidth]{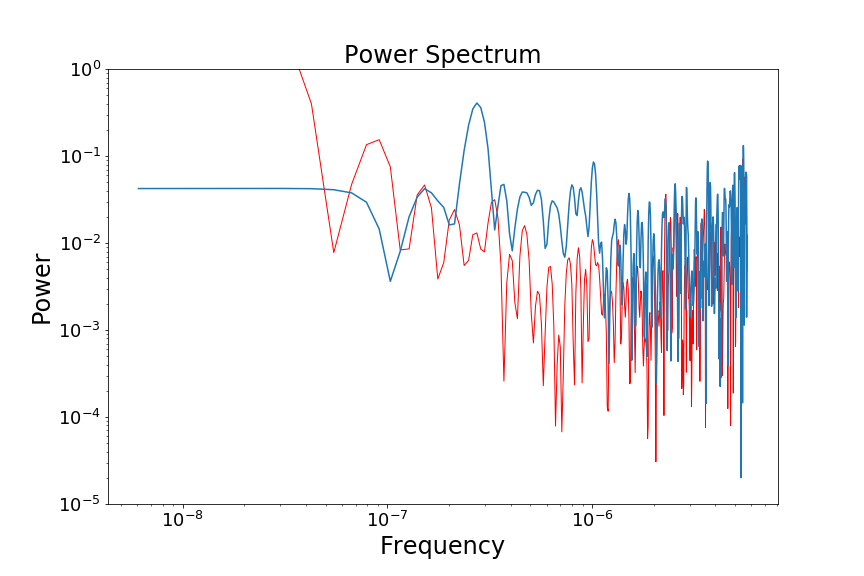}
%
   \caption{(left) Power Spectrum for NGC 4945, MJD 54003-54193, 2-10 keV. The Lomb-Scargle periodogram is shown based on the $\it{RXTE}$ observations of NGC 4945 between MJD 54003-54193 in the 2-10 keV bandpass. A QPO is located at 0.274 $\mu$Hz, corresponding to a period of 42.3 $\pm$ 3 days. A false alarm probability \citep{1986ApJ...302..757H} was determined to be $< 3 \%$. The window LSP in red shows no peak at 0.274 $\mu$Hz.}
\end{figure}

\section{Data Analysis}
\subsection{RXTE Data Analysis}
To analyze the power spectrum of NGC 4945, the $\it{RXTE}$ AGN light curve files were run through a program to determine the Lomb-Scargle periodogram (LSP). The input data is loaded into three arrays:
\setlist{nolistsep} 
\begin{enumerate}[noitemsep]
\item x (the MJD midpoint of the observation),
\item y (the flux in the selected band),
\item dy (the uncertainty in the flux).
\end{enumerate}
The program outputs the light curve and the LSP. The finite frequency grid spacing near the QPO causes imprecise estimates of the QPO period ($\sim$2 days in the present case). Using frequency and power values out to the first local minimum on each side of the QPO peak, a Gaussian fit allows a more accurate estimate of the frequency at mid-peak and its uncertainty. 

\begin{figure*}[t]
    \centering
    \includegraphics[scale=3, width=0.49\linewidth]{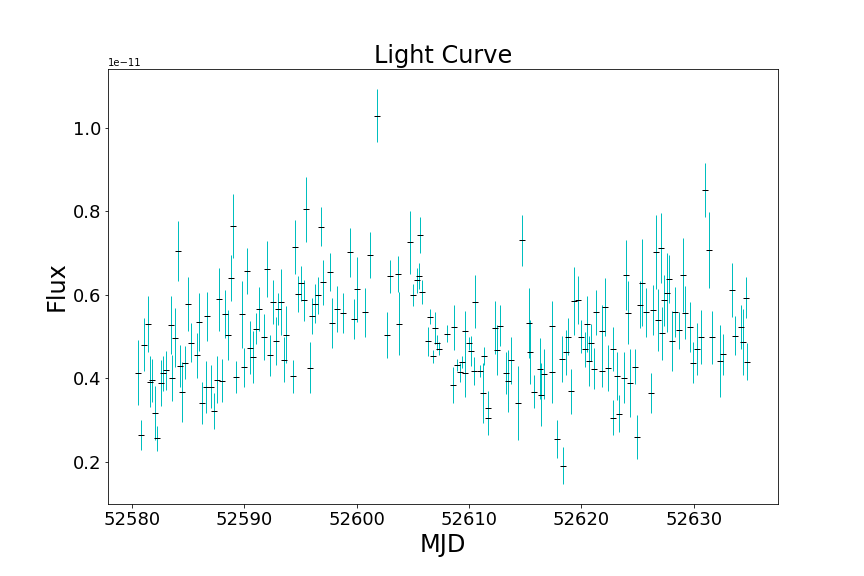}
%
    \includegraphics[scale=3, width=0.49\linewidth]{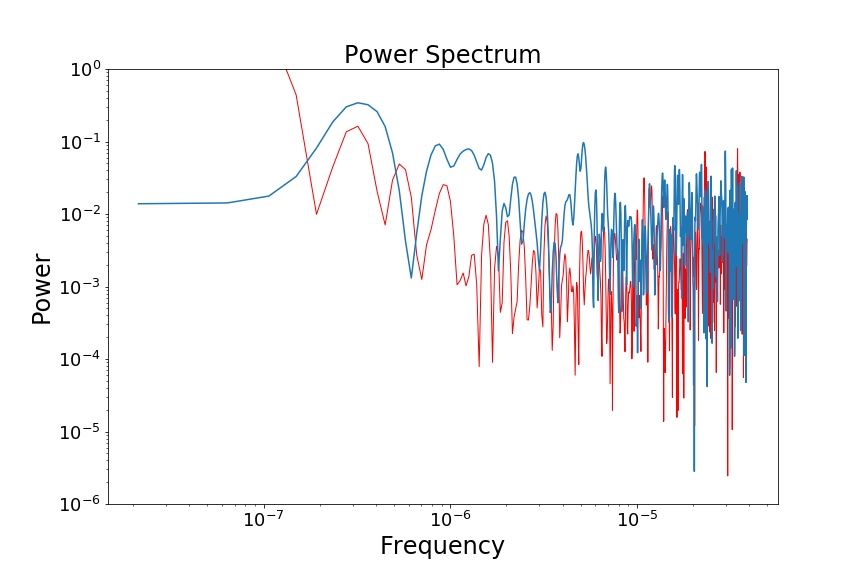}
   \caption{(left) NGC 4945 light curve from $\it{RXTE}$ 2002 data. This was the second longest interval of NGC 4945 data obtained by $\it{RXTE}$. Cycles are clearly visible in the light curve, though the observations cover only $\sim$1.5 cycles.}
    \caption{(right) NGC 4945 power spectrum from $\it{RXTE}$ 2002 data. The Lomb-Scargle periodogram is shown based on the $\it{RXTE}$ observations of NGC 4945 between MJD 54580-52635 in the 2-10 keV bandpass. There is a broad peak close to the value obtained with the longer data arc in 2006-2007. The window LSP is included (in red) and shows a peak close to the QPO. This correspondance between window and data LSP is not in general good. However, there are just 1.5 cycles of the QPO, which is clearly evident in the light curve. The period is near the one reported in the 2007 data.}
    \label{fig:8}
\end{figure*}

Between MJD 54003-54193 ($\it{RXTE}$ proposal 60139) \citep{2001rxte.prop60139M}, the Lomb-Scargle periodogram shows a candidate QPO at 0.274 $\mu$Hz or a period of 42.2 $\pm$ 3 days. Figure 3 shows the NGC 4945 light curve for this time interval. Figure 4 shows the 3-color NGC 4945 light curve for this time interval, where the flux values have been offset as indicated to aid in visibility. Figure 5 shows the power spectrum for NGC 4945 determined from the LSP. The approximately 6-week period appears prominently in the light curve. While a QPO can be seen near this period in the 3 sub-bands, the significance is much lower.

An earlier segment of $\it{RXTE}$ monitoring observations for NGC 4945 (MJD 52580.6 - 52634.8; 2002 November 2 - December 26) contains 183 observations spanning 54 days. During this interval (Figure 6), the spacing between observations varied from a minimum of 1.5 hours (1 RXTE orbit) to a maximum of 1 day. The mean spacing was 0.30 days. While these data span only about 1.5 cycles of the candidate QPO in NGC 4945, the variation again looks cyclic. The LSP returns a QPO (Figure 7) with period 35 days (+9, -6 days; $\sigma$=12.7, FAP = 0.57\%). The additional $\it{RXTE}$ case (from 2002), while not as convincing, does yield a QPO within the quoted uncertainty of the primary result, adding credence to the idea that there is some fundamental time scale in NGC 4945 on the order of 6 weeks. 

\subsection{Quality Assurance}
A useful quality assurance technique is to compute the window transform, or LSP of the window function. This is done using the time series from the actual observations, but with constant flux. This can help rule out false peaks resulting from the data distribution \citep{2018ApJS..236...16V}. We find it useful to plot the window LSP together with the data LSP, allowing easier visual comparison. Applying this technique to our analysis of NGC 4945 helps establish that the $\sim$6-week period does not result from the time spacing of the observations.

To formally establish a False Alarm Probability (FAP), we used the method of Horne $\&$ Baliunas \citep{1986ApJ...302..757H} to estimate the uncertainties in the QPO frequency. For detecting a signal in a power spectrum, we must have a good idea of the contribution to the power spectrum from the noise. We can use this to calculate the probability that a value in the power spectrum could be attained solely by chance. If this probability turns out to be low, then the QPO detection is likely to be real. FAP can be defined as: 
\begin{equation}
FAP(z) \approx  1 - \left[1 - e^{-z}\right]^{N_i}. \label{FAP}
\end{equation}
The values of z are the significance of the maximum power values, normalized by their total variance. The  number of independent frequencies ($N_i$) were related empirically to the total number of frequencies ($N_0$) by: 
\begin{equation}
N_i = -6.362 + 1.193 N_0 + 0.00098 N_0^2. \label{N_i}
\end{equation}
The total number of frequency values in the periodogram is determined by the choices of upper and lower bounds, together with the grid spacing. 

For our principal result, the QPO is detected at significance 10.2-$\sigma$, $N_i$ is 773, based on $N_0$=471. The FAP was 2.87\%. We looked for QPOs in all NGC 4945 data ($\sigma$=7, FAP=99.99\%) and the full 2006-2007 segment ($\sigma$=6, FAP=98\%). Fluctuations looked much more periodic in the second half of the interval, so we made many additional runs of intermediate span. We found the best fit (highest $\sigma$, lowest FAP) for MJD 54003-54193. We have similar experience with the analysis of other AGN in the RXTE archive. Where there are noticeable gaps, we usually need to skip data from the beginning, end, or both, to obtain a light curve that is well sampled. Visual inspection of the light curve often leads to the consideration of shorter intervals that yield improved LSP fits. \textbf{In the present case, we believe that the QPO may only be stable for the 190-day period.}  

\section{Discussion}
\subsection{Use of Unevenly-Sampled Data}
A 2005 study \citep{2005MNRAS.362..235V} questioned whether AGN QPOs would be detectable with data then available. In one simulation, a QPO from a BHXRB was scaled to a typical AGN mass and superimposed on the sampling pattern of the best-sampled AGN in the RXTE archive. The scaled QPO was not detected. A typical $\it{RXTE}$ AGN monitoring campaign might have an observation length of 2 ks, with an observation spacing of 2 days, for a live observation percentage of about 1\%. Also, signal-to-noise (S/N) does not appear to be an issue, based on the average percent errors of the observations. The data span for our primary result covers about 6 cycles of the candidate QPO. While it \textbf{would be preferable to have a much larger number of cycles present \citep{2005MNRAS.362..235V}, this will not always be possible if the QPO is intermittent.} Compared to a typical BHXRB, a typical AGN has about $10^3$ less X-ray flux, but is $10^5$ times more massive, so a corresponding QPO would have a period $10^5$ times longer. Thus, the AGN emits about $10^2$ more photons over its QPO cycle than the BHXRB emits over its QPO cycle. If detected, the AGN QPO may allow resolution of the QPO waveform in the time domain with far greater precision than is possible for a BHXRB.

For uneven spacing, computing a Nyquist-like limit for the LSP involves finding the largest factor p, such that each spacing $\Delta t_i=t_{i+1}-t_i=np$, an exact integer multiple of the factor. For large numbers of observations, the factor p becomes infinitesimally small \citep{2018ApJS..236...16V}, but an upper limit f=0.5/$\Delta$ exists, where $\Delta$ is the limiting accuracy to which time is recorded \citep{2006MNRAS.371.1390K}. The observed and true Fourier transforms satisfy
\begin{equation}
\hat F_{obs} = \hat F_{true} * \hat W,  \label{What}
\end{equation}
where $*$ is the convolution operator, and $\hat W$ is the transform of the window function, 
\begin{equation}
W(t) = \sum_i \delta(t-t_i).  \label{W}
\end{equation}
For $t_i$ distributed evenly, $\hat W$ has a minimum at f = $0.5/\Delta t$, the classic Nyquist frequency, and a spike at f = $1/ \Delta t$. But for uneven-sampling, $\hat W$ is much more complicated, and typically has large amplitude out to frequencies $\sim$$1/ \Delta t$. This results in a lot of power-mixing, making it hard to cleanly measure the Fourier amplitude at any frequency (Krolik, private communication).

In our LSPs, we see the predicted noisiness at high frequencies, but see clean features at low frequencies, far from any Nyquist limit. These detections are limited by the data span or lack of stationarity in the QPO. A possible physical explanation for sporadic QPOs could be a halting of the precession of the X-ray emitting region \citep{1998ASPC..138...75C}. Lense-Thirring precession ceases if the inclination between the accretion disk and the spinning BH becomes 0. Through the Bardeen-Petterson effect \citep{1975ApJ...195L..65B}, friction herds gas into the equatorial plane of the BH, driving the inclination downward. QPOs would occur during transitional or unstable periods, at times when the inner disk experiences major changes.

\subsection{Implications of 42-day Period}
While smoothly-varying broad peaks in the low-frequency portion of the periodogram can reflect the entire time window or harmonics thereof, this LFQPO looks significant after checking for these cases. It does not match other periodicities of the $\it{RXTE}$ orbit like the $\sim$53-day nodal regression cycle. A linear trend in the envelope of the power spectrum (with frequency axis on a log-scale), shows a power-law dependence of power versus frequency in the power spectrum. Many cases show breaks in the power law ("knees" or "elbows"), features that have been extensively studied in $\it{RXTE}$ AGN light curves \citep{2005ApJ...635..180M}. Although we did not attempt to formally fit for a knee or break frequency in the smoothly broken power law, figure 5 appears to generally agree with earlier published break frequency values of $\sim$1 $\mu$Hz \citep{2004AIPC..714..190M}. 

NGC 4945 contains an $H_{2}$O maser which has been used to constrain the SMBH mass to about 1.4 $\times 10^6 M_{\odot}$ \citep{1997ApJ...481L..23G}. NGC 4945 is one of the brightest Seyferts at 100 keV, but is largely absorbed below 10 keV with an absorption column of 4.5 $\times$ $10^{24}cm^{-2}$ \citep{2000ApJ...535L..87M}. With total bolometric luminosity of $\sim$2 $\times$ $10^{43}$erg/s, the source is radiating at $\sim$10$\%$ of the Eddington luminosity. Circumnuclear gas has been modeled as a reflector at a distance $\geq$ 30 pc, and $\it{Chandra}$ images revealed a resolved, flattened, $\sim$150-pc-long clumpy structure, with a spectrum consistent with cold reflection of the AGN emission \citep{2012MNRAS.423L...6M}. $\it{NuStar}$ observed NGC 4945 three times in 2013 for about 50 ks each time. These data were used \citep{2014ApJ...793...26P} to analyze the power spectrum in the spectral range 10-79 keV, but only for frequencies above $\sim$1$0^{-5}$ Hz.

\subsubsection{Keplerian Orbital Period}
The detection of a $\sim$42-day QPO in NGC 4945 scales to $\sim$26 seconds for a mass of 10 $M_{\odot}$. This is outside of the range where QPOs have generally been observed in BHXRB. Scaling arguments relating QPOs between AGN and BHXRB often assume 10 $M_{\odot}$ for the BHXRB mass. However, BH spin can have a large effect on the BHXRB mass determination \citep{2015PhRvD..91l4030J}. If the $\sim$42-day period is a keplerian orbital frequency, then the hot spot would be at nearly 1000 $R_{S}$, considering the well-constrained NGC 4945 mass. Such material would be \textbf{significantly cooler (T $\approx$ $T_{ISCO}$/100)} and not expected to emit X-rays, so we consider other mechanisms. 

The orbital timescale can be parameterized \citep{1999ApJ...514..682E} as: 
\begin{equation}
t_{orb} \approx 0.33 M_7 \left(\frac{r}{10\,R_{S}}\right)^{1.5} \text{days}.  \label{1}
\end{equation}
$M_7$ is the black hole mass in umits of $10^7M_{\odot}$, r is the distance from the BH system barycenter, and $R_S \equiv \frac{2GM}{c^2}$ is the Schwarzschild radius.

This can be inverted to solve for r:
\begin{equation}
r \approx 21 \left[\frac{P(\text{days})}{M_7}\right]^{2/3} R_S.  \label{2}
\end{equation}
The value of r $\approx$ 945 $R_S$ could also be the separation of a binary SMBH. Such models have been proposed \citep{1996ApJ...460..207L} and while we don't rule out this explanation, we see no other compelling supporting evidence.

\subsubsection{Lense-Thirring Precession Period}
The Lense-Thirring precession period is:
\begin{equation}
P_{LT} = \frac{8 \pi GM}{c^3a} \left(\frac{r}{R_S}\right)^3.   \label{P_LT}
\end{equation}
Parameterizing with $P_{LT}$ in days and black hole mass in $10^7$ \(M_\odot\) ($M_7$), the radius of the hot spot becomes:
\begin{equation}
\frac{r}{R_S} = \sqrt[3]{\frac{P_{LT}c^3a}{8\pi GM}} = 4.117 \;\sqrt[3]{\frac{P_{LT}(\text{days}) \;a}{M_7}}.   \label{param}
\end{equation}
So, for this QPO at 42.27 days, and $M_7$ = 0.14 for NGC 4945, the hot spot radius is 27 $r_s$ for a dimensionless spin parameter of $a \approx 1$. The SMBH spin for NGC 4945 does not appear to be well constrained. The radius would be 22 $r_s$ for $a = 0.5$, or 13 $r_s$ for $a = 0.1$. Based on the quoted uncertainty of $\pm$ 3 days, the radial full width of the feature would be $\sim$5$\%$ of that radius.

Even in the optical band, QPO searches in AGN are bearing fruit. Using data from the $\it{Kepler}$ mission (whose primary mission was the search for exoplanets), a peak was found in the power spectrum of the optical light curve of KIC 9650712 (a narrow-line Seyfert 1 galaxy). The QPO is reported at log $\nu$ = -6.58 Hz, corresponding to a period of 44 days \citep{2018ApJ...860L..10S}. For the quoted mass (logM = 8.17), this feature would be at $\sim$6 $R_S$ if caused by Lense-Thirring precession. A 5-day feature in the $\it{Kepler}$ light curve of Zw 229-15 \citep{2014ApJ...795....2E} could also be identified with Lense-Thirring precession. For $M_7 \approx 1$, this feature would be at $\sim$7 $R_S$. Timing features in blazars have also been investigated using combined $\it{Fermi}$ ($\gamma$-ray) and optical data \citep{2016AJ....151...54S}. 

We acknowledge that deriving the radius using point-mass assumptions is an oversimplification of the true physics. An extension of the model replaces the test mass with the entire accretion flow inside the disk truncation radius \citep{2012MNRAS.419.2369I}, such that the entire hot flow precesses as a solid body with frequency given by a surface-density-weighted average of the $f_{LT}(r)$.

\subsubsection{Other Possibilities}
\textbf{Both Lense-Thirring and Keplerian frequencies are used to explain QPOs in} the relativistic precession model \citep{1998ApJ...492L..59S}. Many other models exist in the literature, such as parametric resonance instabilities \citep{1997ApJ...477L..91N}; accretion ejection instability \citep{1999A&A...349.1003T}; propagating oscillatory shock \citep{1996ApJ...457..805M}; two-oscillator model \citep{1999ApJ...518L..95T}; beat frequency model \citep{1998ApJ...508..791M}; diskoseismic or corrugation models \citep{2009ApJ...692..869R,2011ApJ...736..107O}; MHD loop \citep{2009MNRAS.392..264S}; precessing inner flow \citep{2009MNRAS.397L.101I}; inner flow / truncated disk \citep{2010MNRAS.404..738C}; toroidal Alfv\a'{e}n wave oscillations \citep{2012RAA....12..661W}.

We have concentrated on Keplerian and Lense-Thirring frequencies as these are common elements of many models. Other interpretations are possible. The magnetic Alfv\a'{e}n wave model has a substantial contribution to the QPO frequency determined by the disk thickness to radius ratio.

Several ongoing missions ($\it{XMM}$-$\it{Newton}$, $\it{Chandra}$, $\it{NuStar}$) perform long-stare observations that have facilitated the search for HFQPOs in AGN. On the other hand, the RXTE AGN archive offers long, but sparse monitoring campaigns that offer the chance to find LFQPOs in AGN. We suggest that further work on additional sources in the RXTE archive is warranted. Study of the QPO phenomenon in both AGN and BHXRB will contribute to a better understanding of each.

\acknowledgments
We thank Greg Madejski, Jean Swank, Julian Krolik and especially Alex Markowitz for valuable and informative discussions. We thank our anonymous referee for many insightful comments which helped to improve the quality of this paper.

This work utilized light curves provided by the University of California, San Diego, Center for Astrophysics and Space Sciences, X-ray Group (R.E. Rothschild, A.G. Markowitz, E.S. Rivers, and B.A. McKim), obtained at \url{http://cass.ucsd.edu/~rxteagn/}.

\bibliography{QPO}

\end{document}